\documentclass[12pt]{article}
\usepackage{graphicx}

\usepackage{cite}
\usepackage{amsmath}

\begin{document}

\title{SOCIOPHYSICS: A NEW APPROACH OF SOCIOLOGICAL COLLECTIVE BEHAVIOUR. I. MEAN-BEHAVIOUR DESCRIPTION OF A STRIKE \\ $\,$}

\makeatletter
\renewcommand{\maketitle}{\bgroup\setlength{\parindent}{0pt}
\begin{flushleft}
  \textbf{\@title}

  \@author
\end{flushleft}\egroup
}
\makeatother

\author{SERGE GALAM, YUVAL GEFEN (FEIGENBLAT) and \\ YONATHAN SHAPIR\\ 
Department of Physics and Astronomy \\
Tel-Aviv University \\ Ramat-Aviv, Israel}

\maketitle

$\,$\\ \textbf{SUBMITTED IN {\large 1980}, PUBLISHED IN {\large 1982}}\\ 
Journal of Mathematical Sociology, Vol. 9, pp. 1-13\\ \\
$\,$\\ ``TO WORK OR NOT TO WORK, THIS IS THE QUESTION" \\(Anonymous)\\ \\
A new approach to the understanding of sociological collective behaviour, based on the framework of critical phenomena in physics, is presented. The first step consists of constructing a simple mean-behaviour model and applying it to a strike process in a plant. The model comprises only a limited number of parameters characteristic of the plant considered and of the society. A dissatisfaction function is introduced with a basic principle stating that the stable state of the plant is a state, which minimizes this function. It is found that the plant can be in one of two phases: the "collective phase" and the "individual phase". These two phases are separated by a critical point, in the neighbourhood of which the system is very sensitive to small changes in the parameters. The collective phase includes a region of parameters for which the system has two possible states: a "work state" and a "strike state". The actual state of the system depends on the parameters and on the "history of the system". The irreversibility of the transition between these two states indicates the existence of metastable states. For these particular states, the effect of small groups of workers or of a small perturbation in the system results in drastic changes in the state of the plant. Other non-trivial implications of the model, as well as possible extensions and refinements of the approach, are discussed. 

\section{INTRODUCTION}

The strict separation between different disciplines of human knowledge, imposed arbitrarily for psychological or sociological reasons, has been definitively suppressed by the internal and independent development of each discipline. Yet, although nowadays the disciplines of psychology, sociology and biology are somewhat linked, physics still remains essentially isolated from all that is concerned with life, and in particular, with man. 

Physics was successfully applied in some branches of biology [such as molecular biology (Marois, 1969)], nevertheless there still exists opposition to the extension of physics into other domains. One apt example is the discipline of sociology. There is, however, one aspect of human behaviour which is directly related to physics; namely, collective behaviour. In physics this term refers to situations where the behaviour of each small constituent of a large system is correlated with the behaviour of others, and all components of the system lose their "individual" character. The importance of the collective factor in human society need hardly be stressed, since social phenomena on a macroscopic level are manifested in various situations, e.g., demonstrations, popular movements, revolutions, etc. The substantial progress achieved in the study of collective phenomena in physics, on one hand, and the apparent similarity between these phenomena and the respective sociological phenomena, on the other hand, are a temptation towards the employment of a physical approach for the purpose of describing some types of social behaviour (Voronel, 1974; Callen and Shapiro, 1974, Weidlich, 1980). 

Yet, a physical approach may be criticized for a variety of reasons, on the grounds of an essential difference between an individual and a particle. These objections may be summed up in the following three arguments; 

(1) A human being is a very complex system consisting, among other things, of feelings, motivations... and possibly, a free will. Compared to physical systems (atoms, particles, etc.), human complexity is by far greater. 

(2) Moreover, individuals differ from one another in their character, behaviour, etc., while in physics particulars of certain entities are identical. 

(3) Owing to the complexity of human society, a multitude number of factors is required to describe its properties, as compared to the relatively few parameters needed in the case of physical systems. 

These objections, however, are overruled by the following argumentation: Were we to take the analogy with physics a step further, we would observe that multi-factored
macroscopic physical systems can be adequately described by the use of a small number of parameters. In the physics of large systems [thermodynamics (Landau and Lifshitz, 1980; Callen, 1960)], we are very rarely interested in describing a situation by using a great number of parameters; although, theoretically it might furnish us with a precise description, it is redundant and ineffective. We, therefore, use a few macroscopic parameters (in addition to some microscopic characteristics of the system) for the purpose of obtaining a realistic description of our ensemble. 

We thus see that the total set of individual aspects of each component of a large system is not necessarily relevant. It is noteworthy that physics is capable of dealing with different behaviours of identical particles and of considering systems composed of different particles (see for example, Kirkpatrick,1979). In many cases, the introduction of refinements of that kind does not change the main features of collective phenomena [sometimes differences among particles induce more complex collective behaviour. Recently, physical techniques have been developed for the study of such situations (Kirkpatrick, 1979 and Wilson and Kogut, 1974)]. 

We intend, however, to restrict ourselves in the first article to identical individuals. Subtler models will be constructed in subsequent papers. One central concept in the thermodynamics of physical systems is universality (Wilson and Kogut,1974): there are situations of great interest (critical collective phenomena) where different systems exhibit essentially similar behaviours. In such cases, many of the differences among given systems are irrelevant and we designate them as systems belonging to the same universality class. This inspires us with the hope that we will be able to relate social ensembles to one or another well-known universality class.

In fact, those who wish to protect the concept of individuality,  A theory explaining collective behaviour will enable us to single out collective factors of human behaviour, and will shed a light on individual aspects as well. The domain of gravitation illustrates our point rather well; only after the existence of gravity was recognized and its laws understood, the road was cleared for overcoming the limitations it imposed on man.

In the present paper we study a simple model of collective behaviour,  which we propose to call a model of mean-behaviour (M.B.). This model is based on the Mean Field Theory of Weiss (1907). We apply this model to the description of a strike process in a big plant. We consider an ensemble of identical individuals, taking into account the contacts between them and a few sociological characteristics of the society they belong to. In order to find what is the actual state of the collective out of all the a priori possible states, we introduce the principle of minimum dissatisfaction. This is a primary postulate in our theory.

The following part of the present paper is concerned with the description of the model, definitions of the parameters involved and formulation of the principle of minimum dissatisfaction. The third part contains an analysis of the model. In the last part we outline the main conclusions, and possible directions of future study.

\section{THE M.B. MODEL: DEFINITIONS AND FORMULATION}

It follows from the introduction that although sociological problems contain inherently a multitude number of factors,  in the M.B. model we shall employ only a few parameters, 
hoping to obtain a non-trivial realistic description of social behaviour. It should be emphasized that the actual set of parameters used for the description of the problem is in no way unique.

In the present simplified model, which is only a step in the new direction of sociological analysis proposed here, we neglect many factors that might be relevant to a complete and more realistic description.

\underline{Realization of the Model}: In order to illustrate the M.B. approach, a situation of a strike in a large plant has been chosen. For the moment, we disregard the differences among the workers' occupations in the plant, their specific production potential, their socioeconomic background, their personality, etc. In other words, we consider the N workers (N is a large integer) as identical (that doesn't mean that they necessarily behave identically).

a. \underline{Microscopic description}. The first step in constructing the model concerns the description of individuals in the ensemble.

{\it Individual production} $\tilde \mu_i$ is a measure of the amount of production of the $i^{th}$ individual; $\tilde \mu_i$ is bounded between 0 (the individual is not working) and 1 (the individual is working at maximal production).

{\it Normalized individual production}, , is related to by the expression:
\begin{equation}
\mu_i=2(\tilde \mu_i- \frac{1}{2}) ,
\label{1}
\end{equation}
i.e.:

$\mu_i=-1$ means that the individual is striking,

$\mu_i=+1$   means that the individual has reached his maximal production.

{\it Contact} $J_{ij}$ is a measure of the mutual interaction between individuals. Sampling two individuals, i and j, out of the ensemble, the interaction between them will induce some amount of dissatisfaction, which can be written as\footnote{Here it was assumed that $J_{ij}$ is a symmetric tensor: $J_{ij}=J_{ji}$. Otherwise, expression (2) should be modified to: $- \frac{1}{2}J_{ij}\mu_i \mu_j -\frac{1}{2}J_{ji}\mu_i \mu_j$
}:
\begin{equation}
-J_{ij} \mu_i \mu_j .
\label{2}
\end{equation}

 $J_{ij}$ is a positive parameter (and in the subsequent analysis will be taken to be some constant $J_{ij}\equiv J$ , i.e. independent of the indices i and j). To put it more explicitly: if the 7th worker is fully working $(\mu_7 = +1$ and the 168th worker is striking $(\mu_{178} = -1$) then the amount of dissatisfaction due to their mutual interaction only is  $J_{7, 168}$ and is greater as $J$ becomes larger. Had they been in the same "work state" (e.g.  $\mu_7=\mu_{178} = +1$) the dissatisfaction would have been lower (in this example:  $-J_{7, 168}$).

$J_{ij}$ depends, among other things, on internal conditions at the plant; e.g., if the workers spend their breaks and vacations together or share common activities, we would expect J to be larger.

b. \underline{Macroscopic description}. We are now interested in quantities concerning the ensemble as a whole.

{\it Normalized mean production} is obtained by averaging over $\mu_i $:
\begin{equation}
M= <\mu_i> = \frac{1}{N} \sum_i \mu_i .
\label{3}
\end{equation}
M describes the "work state" of the plant. By definition it is bounded between -1 and +1. M is a measure of the intensity either of working in the plant (if M is positive) or of striking (if M is negative).

{\it Entropy} S: It should be clear that for any given mean production M, there is, in general, more than one microscopic configuration (characterized by the values of $\{\mu_i$\}), which corresponds to M (i.e., averaging over the $\{\mu_i$\} yields the same value of M). Let us denote by L the number of different configurations which have the same average. The entropy is then defined by:
\begin{equation}
S=log L .
\label{4}
\end{equation}

In case L is large (high entropy), knowledge of M will not provide us with much information about what happens in our ensemble microscopically. The opposite is true for a small L. This definition of S coincides with the concept of entropy in other domains (see, for example, Kirkpatrick, 1979).

{\it Social permeability} 1/T measures the potential response of an individual to some influence due to the social atmosphere he is embedded in.

{\it Effective salary} H: This includes the worker's salary as well as seniority rules, pension and vacation rights, etc. (We do not propose here a quantitative criterion of H.) We assume that there is some range where the contribution to the dissatisfaction is proportional to:
\begin{equation}
- \mu_i H .
\label{5}
\end{equation}

The linear form of Equation (5) is not essential for the forthcoming analysis, and comes only for reasons of simplicity. Notice that since the dissatisfaction function F (discussed below) may contain other (possibly competing) terms, the total dissatisfaction need not necessarily be proportional to either $\mu$ or H.

c. \underline{Basic principle}. In order to analyze the different situations in our model, we need a guiding principle. Once the values of the parameters of the problem are given, this principle will enable us to know what the actual state of the system is. Therefore, we introduce the {\it Principle of Minimum Dissatisfaction} (PMD): We define F, which is a function of H, T and J. The PMD would then state:
\begin{equation}
\begin{split}
&\text {Given H, T and J, the variables of the system M and S assume values}   \\ & \text {$M_0$ and $S_0$ (both functions of H, T and J) such that F is minimum.}
\end{split}
\label{6}
\end{equation}

More precisely, this principle states that for a given contact J, salary H and permeability T, the system as a whole will tend to be in its most "satisfied" state (or will reach its most harmonious state). This state is described by the values of the production $M_0$ and the entropy $S_0$ (which are determined using the PMD).

A general expression for the differential of F is (for a fixed J);
\begin{equation}
dF(H,T)=-MdH-SdT .
\label{7}
\end{equation}
The derivation of this expression is given in the Appendix. 

Table 1 contains a summary of the variables and the parameters defined in this section.

\begin{table}
\begin{center}
\begin{tabular}{ lcc } 
 \hline
 Quantity & Notation & Comment  \\ [1.0ex]
 \hline 
\underline{Microscopic Level} &  & \\
Individual production &  $\tilde \mu_i$ &  $0\leq \tilde \mu_i\leq 1$\\
Normalized individual production & $\mu_i$ &  $-1\leq \mu_i\leq 1$ \\ 
Contact & $J_{ij}$ & $J_{ij}=J>0$ \\[1.0ex] 
\underline{Macroscopic Level}& & \\
Number of workers &$N$ & $N >1$ \\ 
Entropy & $S=log L$& \\
Social permeability & $\frac{1}{T}$& \\ 
Mean production  &$M=<\mu_i>$  & $-1\leq M \leq 1$\\
Effective salary & H & \\
Dissatisfaction function &F & Principle of Minimum Dissatisfaction\\ 
\hline
\end{tabular}
\caption{Definition of parameters and variables of the MB Model.}
\label{t1}
\end{center}
\end{table}

\section{ANALYSIS OF THE MODEL}

a. \underline{Existence of two phases}. We consider first the dissatisfaction caused by the interactions among the individuals and the influence of H. Employing (2) and (5) we can write this quantity as;
\begin{equation}
-\sum_{i,j} J_{ij} \mu_i \mu_j -H \sum_i \mu_i  \xrightarrow[\text{$\{J_{ij}\}=J$}] {\text{\ }}  -\sum_{i,j} J \mu_i \mu_j -H \sum_i \mu_i
\label{8}
\end{equation}

A simple way to look at the problem is to observe that every individual is affected by H and the other individuals. As a reasonable approximation we can substitute the effect of the above factors by some quantity multiplying each . Each individual is contributing to this quantity. That is, the state of an individual is determined by the other individuals and influences them at the same time. Thus, the expression we obtain is a self-consistent equation for the state of the whole ensemble.

Putting it more formally, we can make the approximation of taking averages of quantities appearing in equation (8). We obtain:
\begin{equation}
 -\sum_i \mu_i  \{H+  \sum_j J <\mu_j> \}=  -\sum_i \mu_i  \{ H+  \sum_j J M \} .
\label{9}
\end{equation}
In equation (9) the last term is linearly proportional to M. Thus, the quantity which multiplies each can be written as:
\begin{equation}
\tilde H=H+a M .
\label{10}
\end{equation}
We argue now that M should be proportional $J\tilde H/T$. This argument may be considered as a reasonable phenomenological theory (which is tested by its predictions). In order to understand the inverse proportionality to T, we should recall the fact that H (or $\tilde H$) and T are, in a sense, competing factors.

Thus we may write:
\begin{equation}
M=\frac{C}{T} (H+a M) ,
\label{11}
\end{equation}
(C and a are constants.)

It is clear from (10) that $a \sim T$. Defining:
\begin{equation}
K=\frac{T}{J} ,
\label{12}
\end{equation}
we obtain:
\begin{equation}
M=\frac{C(H/T)}{1-(a'/K)} \text{   with  } a'=\frac{a C}{J} .
\label{13}
\end{equation}

Starting from a large K, we observe that as K decreases, and $H \neq 0$ is kept constant, M increases, until, as K approaches some critical value $K_c= a'$, M diverges.

We can say, that we have two regimes of behaviour for M, or two {\it phases}:
The {\it individual phase} ($K>K_c$) where $M=0$ for $H=0$ (the workers' production distributes somehow around $\mu_i = 0$; in other words, the workers' mean production is $\sim 0.5$).

At $K_c$ something happens: there is a divergency in the quantity $\frac{\partial M}{\partial H}\Bigr\rvert_{K= const.}$, (defined as the differential susceptibility). For $K < K_c$ there is another regime of behaviour. This is the {\it collective phase}, where, as we shall see below, for $H\rightarrow 0, M\neq 0$. In this case, we see that for a zero effective salary, the production of the group of workers is either under or above average: the workers are now striking ($M < 0$) or working ($M > 0$) as a collective.

In order to obtain a more detailed description of the two phases, we apply, in the next subsection, a more detailed phenomenological approach.

b. \underline {Analysis of the phases}. Using a phenomenological approach we select a simple, yet realistic, expression for F, under the constraint that the results already obtained will be included in the present description.
Since at constant T (and K) $\Delta F = - M \Delta H$ (7), the explicit dependence of F on H can be taken as:
\begin{equation}
F (M, H, T) = F (M, H=0, K) -M H .
\label{14}
\end{equation}
We shall proceed by looking for an expression for F(M,0,K). In the vicinity of $K_c$, M, if not zero, is a small quantity. From now on, we can follow the theory of Landau and Lifshitz (1980) and only interpret its results in sociological terms. Within this theory, the system can be found in one of two phases, depending on the value of K. For $K > K_c$ (the individual phase), F has two minima, one of which describes the state of the system. The minimum will be at positive or negative values of M, depending on the effective salary H. For $K < K_c$ (the collective phase), F may have two local minima, $M_0$  and $M_0'$ which are equal only for $H = 0$.

The actual state of the plant (i.e., whether it is a "strike state" or a "work state") depends on its "history" (the phenomenon of hysteresis). Suppose we start from a state of $M > 0$, and the system has $M_0 > 0$ (the plant is in a state of work, cf. Figure 1). In that case, F has two minima - the lowest one represents the actual state of the plant. Decreasing H, the other minimum (with $M_0' < 0$) becomes lower, and the system, being still at $M_0 >0$, is in a metastable state (Figure 2). The fact that the system is at $M_0 > 0$ (though the "strike state" at $M_0' < 0$ is more stable) reflects the dependence of the system on its history. The system may jump to $M_0'$ (i.e., from a "work state" to a "strike state") in one of two cases: either when H decreases further and we reach the limit of metastability (the minimum at $M_0 >0$ disappears (Figure 3) or due to some perturbation. Such a perturbation, either external (e.g., external influence) or internal (small group of working calling for a strike) can virtually raise the dissatisfaction of the workers by at least $\Delta F$ (for a short time) such that the system can "overcome the barrier" and shift to the stable state. A complete "hysteresis loop" is shown in Figure 4.

\begin{figure}[t]
\centering
\includegraphics[width=1\textwidth] {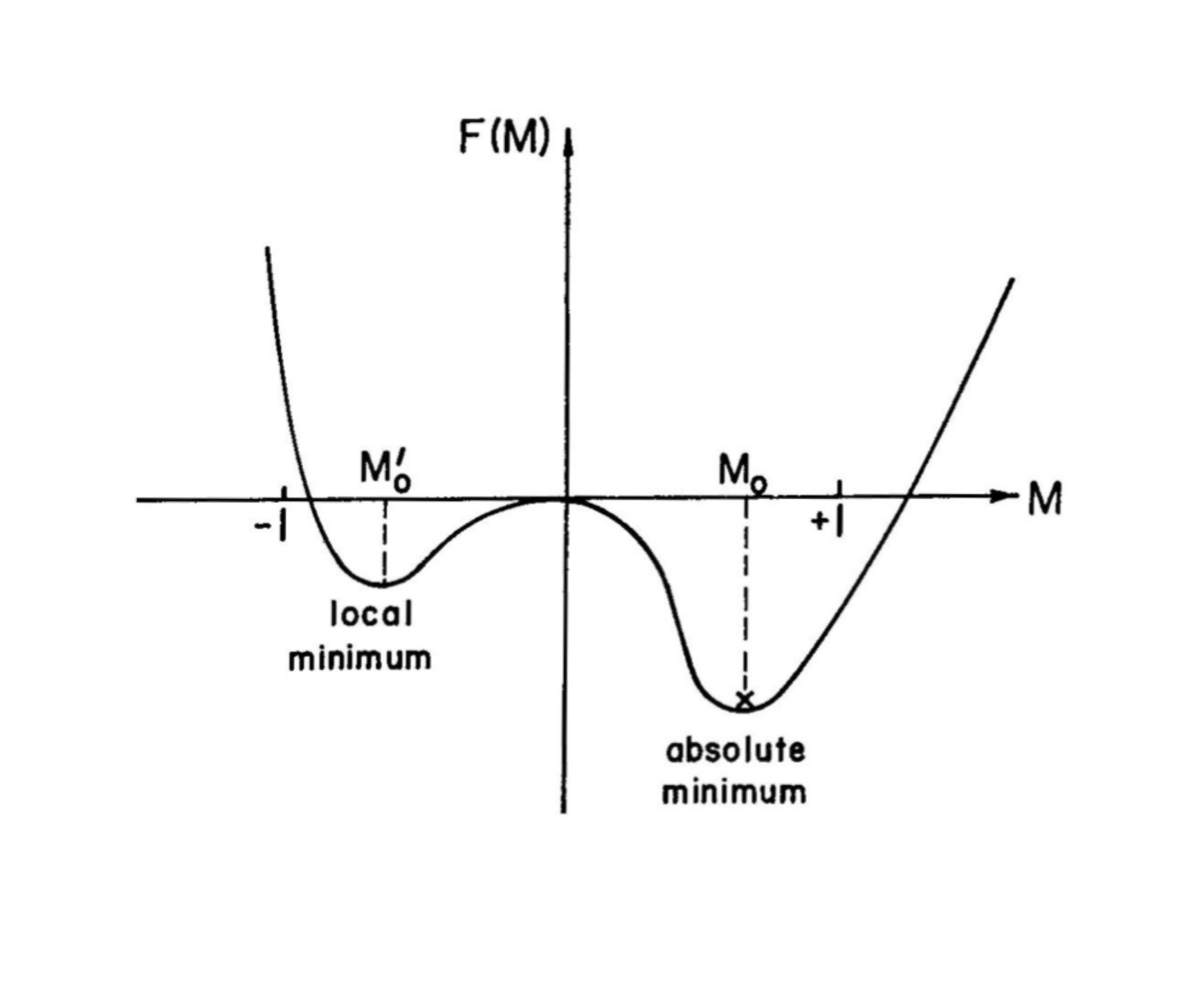}
\caption{Dissatisfaction function F vs. M for $K<K_c$. In the following x denotes the actual state of the system $H > 0$.}
\label{fig1}
\end{figure}  

\begin{figure}[t]
\centering
\includegraphics[width=1\textwidth]{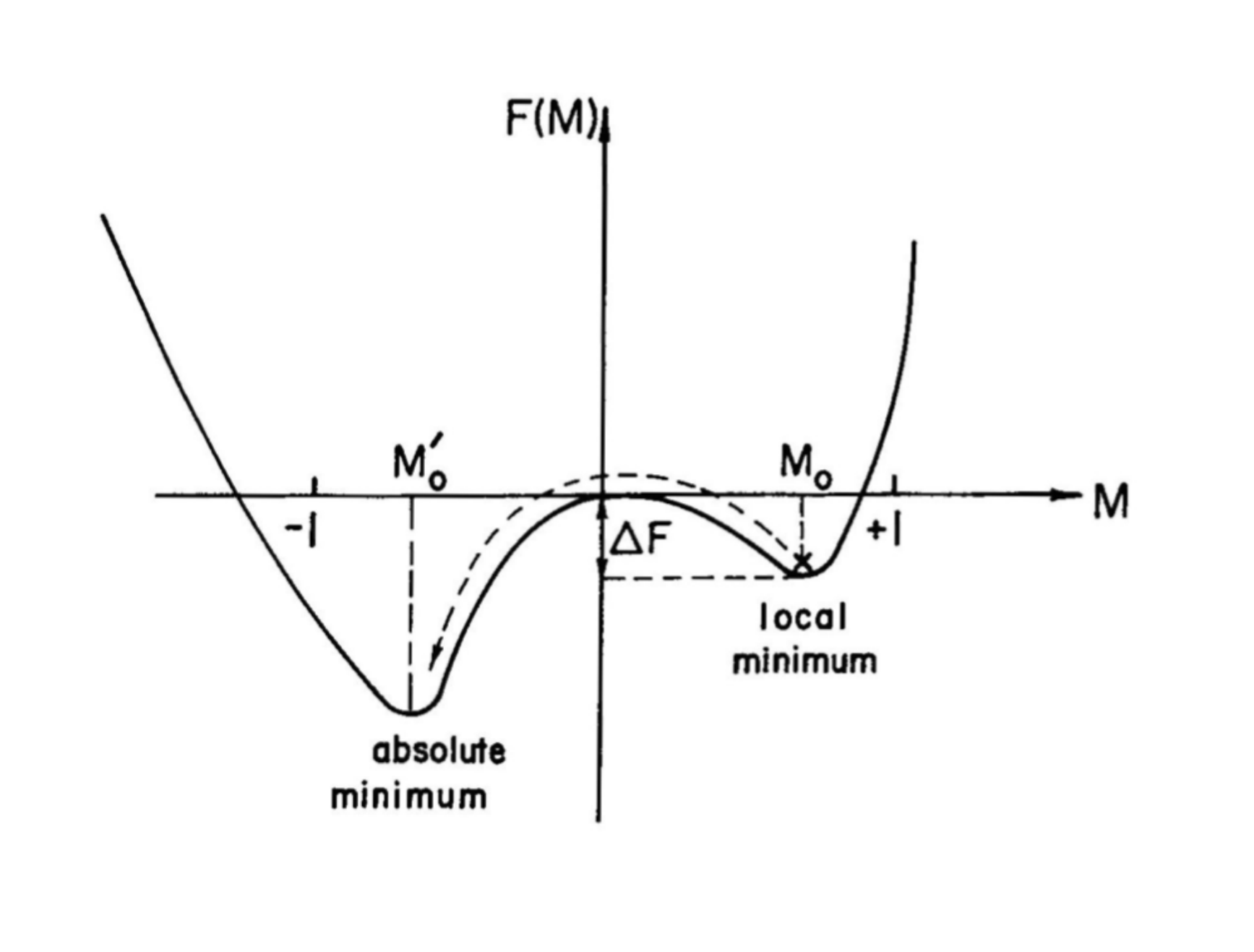}
\caption{A metastable case with $H < 0$. The dashed line denotes a possible jump to a stable strike state due to some perturbation.}
\label{fig2}
\end{figure}  

\begin{figure}[t]
\centering
\includegraphics[width=1\textwidth]{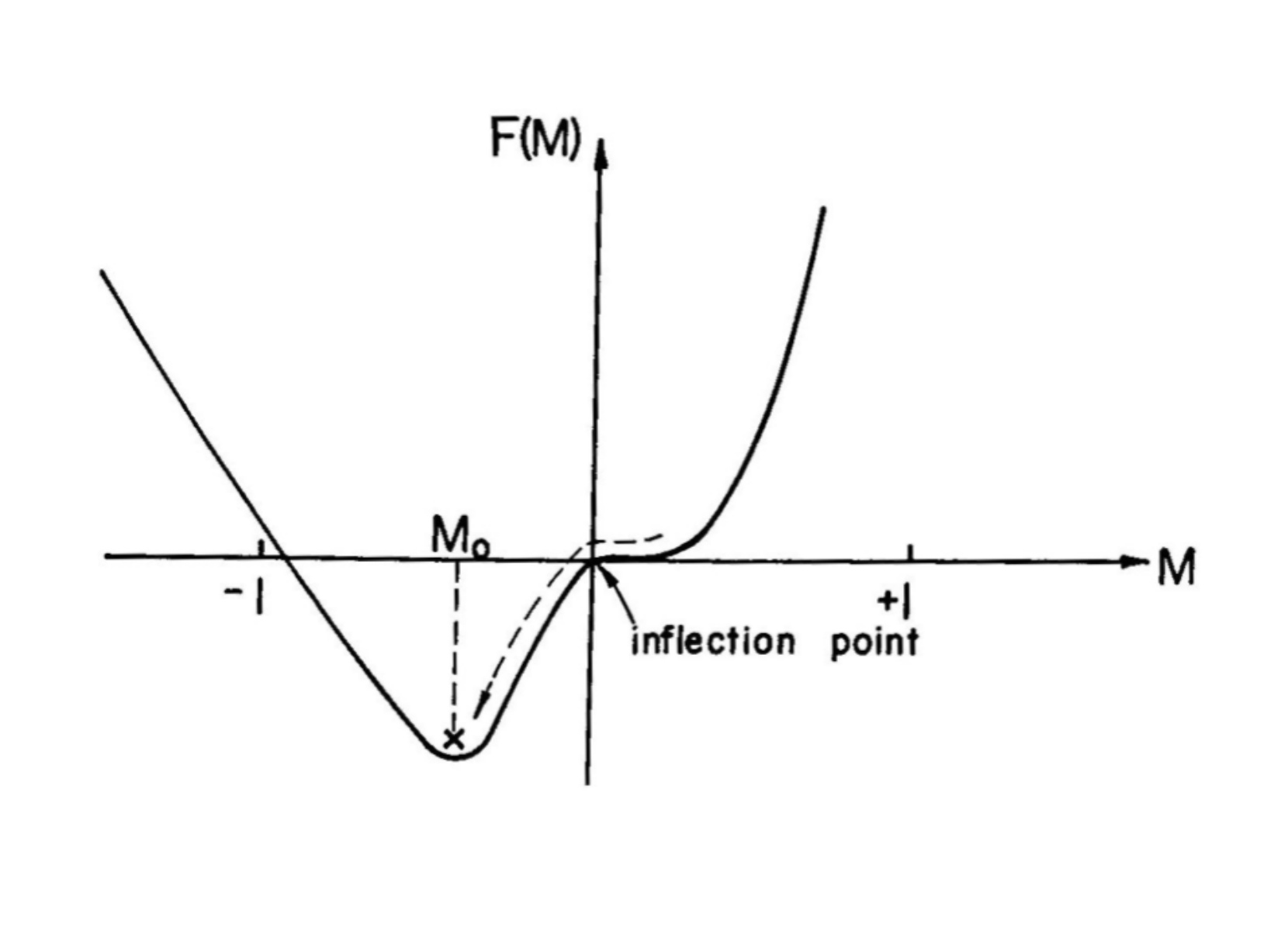}
\caption{The limit of metastability for $H < 0$. F has only one minimum.}
\label{fig3}
\end{figure}  

\begin{figure}[t]
\centering
\includegraphics[width=1\textwidth]{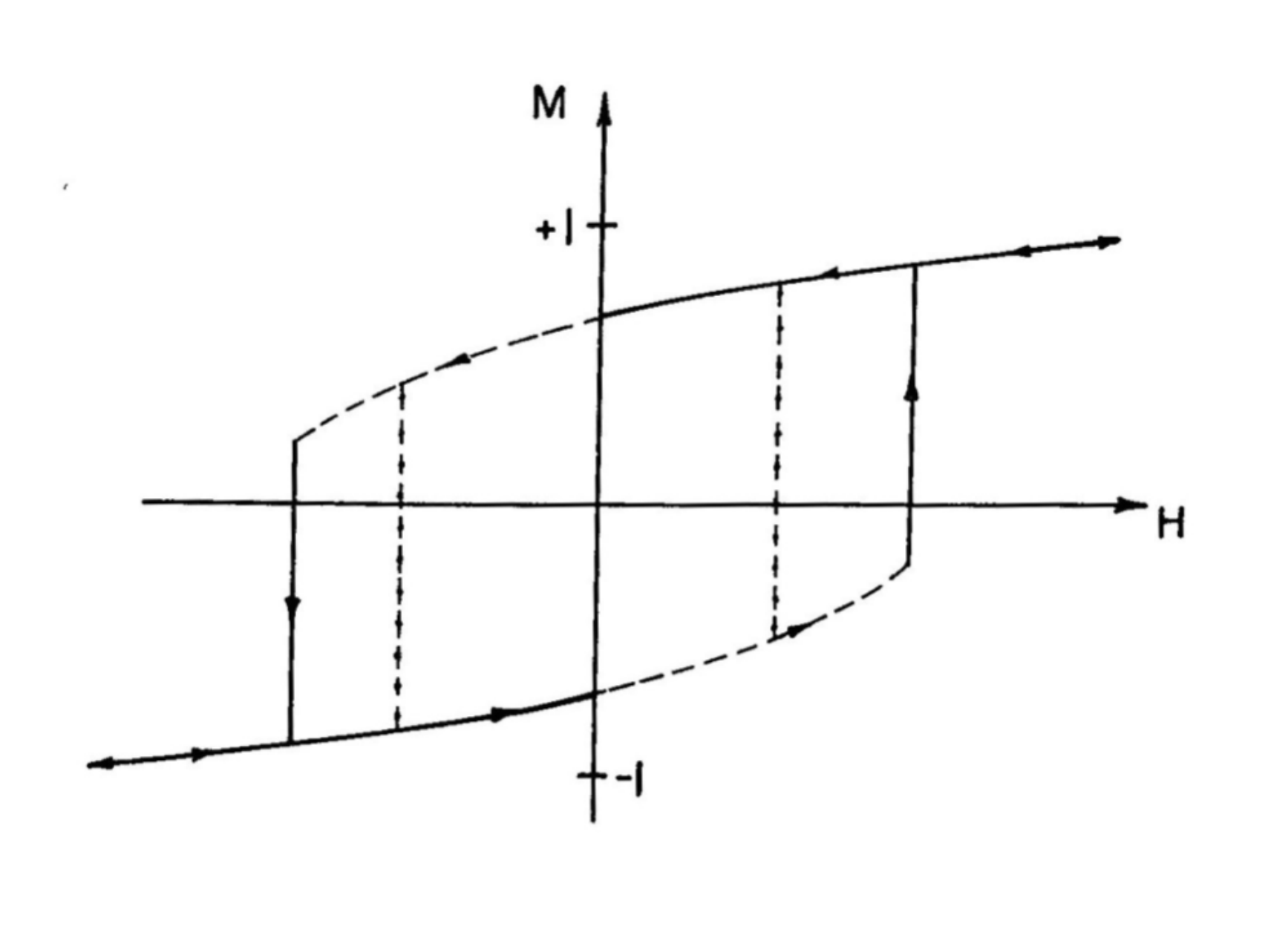}
\caption{M vs. H for $K < K_c$. Arrows show the direction in which H is changed. Solid lines denote stable states. Dashed lines denote metastable states. "Dot-dash" lines show possible jumps from metastable to stable states.}
\label{fig4}
\end{figure}  

We can define the susceptibility of the plant, $\chi$ as, 
\begin{equation}
\chi (K, H)=\Delta M_0 / \Delta H\Bigr\rvert_{K, H} .
\label{15}
\end{equation}
When $K\approx K_c$ and $H \approx 0$ the susceptibility is very large, which means that the workers are very sensitive to small perturbations.

\section{CONCLUSIONS AND DIRECTIONS FOR THE FUTURE}

We begin this section with a brief discussion of the important conclusions derived from our model.

\begin{itemize}
\item 
The ratio between the inverse of the social permeability T and the contact (denoted by $K=\frac{T}{J}$) determined the character of the plant. For K smaller than a certain critical value, the behaviour of the system is essentially "collective." For larger values of K, the character of the system is individual.
\item 
There are metastable states of the system. In those cases the system is in a state of work (apparently a normal, stable state of work), but a small perturbation (e.g., external influence, a small group of workers calling for a strike) may cause the system to shift to a stable state of strike. The same is true for the transition from a state of strike to a state of work. As the metastable state becomes less stable, the perturbation needed is smaller.
\item 
Once a strike breaks out, larger payment is required to drive the workers back from strike to work, than the one required to shift the plant from a metastable to a stable state of work (the hysteresis phenomenon).
\item 
Generally, increasing the salary will result in an increase of the production. But as the intensity of the work (or of the strike) is higher, this change (the susceptibility) is smaller. If the effective salary is small (in absolute value) it is more paying to increase it, especially when K is close to the critical value, than if it is high.
\item 
A plant in a state of work: keeping the salary constant and increasing K results in an increase of the production.
\item 
Consequently, there are two different situations where the plant is very sensitive to small changes in its conditions: a metastable state, and a state with a small effective salary (in absolute value) and K near $K_{critical}$.
\item 
A plant in a state of strike: keeping the salary constant and increasing K results in an increase of the intensity of the strike.
\item 
The transition from a metastable strike state to a stable work state (when the salary is positive) will be effected by a smaller perturbation as K (still smaller than its critical value) increases.
\end{itemize}

It might be useful to indicate some possible generalizations and modifications of our simplified model, making it more applicable to real situations.

The supposition of uniform contact and permeability for all the workers may be reasonable, when the workers form a homogeneous social group. In more complicated situations we may have several groups, each having its own K value. The generalization of our model to this case is straightforward. We may also consider cases where, instead of uniform contact J, we have different values $J_{ij}$ according to some given distribution. It may be interest to go further and deal with the problem where the symmetry $J_{ij}=J_{ji}$ does not hold (the influence of the $i^{th}$ worker on the workers is not equal to the influence of the $j^{th}$ on the $i^{th}$). Due to the fact that the salaries and the needs or expectations of the workers are not equal, it is interesting to study the case with non-uniform H. Another modification is to allow each worker to deviate from the mean behaviour.

In our model, there is symmetry between states of strike and stages of work (every statement including the term "state of work" may be appropriately changed to be a statement about a "state of strike" and vice versa). We may think of other models in which this symmetry is not present.

Some of the above problems may prove to be very interesting, and hopefully, will be analyzed in the future, using physical analogues. We hope that for the more complicated (and realistic) models suggested here, numerical values of the parameters will be experimentally available, thus leading to the applications of theories of this kind to real life.

A final remark is due. In this work, we considered a simplified model. Beyond considering the specific case of work and strike in a plant, we suggest that the importance of this approach lies in the use of physical methods. Starting with very few and simple assumptions, we reach nontrivial results. {\it Without an a-priori assumption on collective behaviour}, we obtain very nonlinear behaviour of the system, including (for a certain region of the parameters) {\it a collective nature as a result}. This is only an initial step. This method may become a well-established approach to the understanding of sociological collective phenomena, provided that a study of further models, as well as an improved translation into the sociologists' vocabulary is carried out.

\section*{ACKNOWLEDGEMENTS}

We would like to acknowledge Prof. A. Voronel for his stimulating support and suggestions. We also thank Prof. E. Ya'ar for fruitful discussions. The assistance of C. Nachmani in carefully reading and editing the manuscript and of Dina Naimark in typing it is gratefully acknowledged.

\section*{REFERENCES}

Callen, H. B. (1960) Thermodynamics. New York: Wiley Toppaz. \\
Callen, E. and Shapiro, D. (1974) Theory of social imitation. Physics Today, July, 23. \\
Kirkpatrick,S.(1979) Models of Disordered Materials. Amsterdam: North Holland.\\
Landau, L. D. and Lifshitz, E. M. (1980) Statistical Mechanics. New York: Pergamon Press. \\
Marois, M. (1969). Theoretical Physics and Biology. Amsterdam: North Holland.\\
Voronel, A. (1974) In Norman A. Chigier and Edward A. Stern (eds.). Collective Phenomena and the Applications of Physics to Other Fields of Science. Brain Research Publications, p. 333. \\
Weidlich, W. (1979) British Journal of Mathematical and Statistical Psychology,  \underline{24} 251; (1980) Collective Phenomena, \underline{3}: 89.\\
Weiss, P. (1907) J. Phys. Radium,  \underline{6}: 667. \\
Wilson, K. G. and Kogut, J. (1974) Phys. Report,  \underline{C12}:75.

\section*{APPENDIX}

In this Appendix we derive an expression for the differential of F. Suppose we somewhat increase T (keeping H and
J constant); the higher the entropy, the greater will be the decrease in the dissatisfaction (because as the disorder is greater, the individual character of each one can be better expressed when we increase T and this results in higher "satisfaction"). Thus, we may expect a contribution of -SdT to dF when we change T to T + dT.

As to the contribution to dF due to the variation of the effective salary from H to H + dH, we see from (5) that the dissatisfaction is changed by $-\mu_i$dH and for the whole system the average change in the dissatisfaction is:
\begin{equation}\tag{A1}
-<\mu_i> dH =- M dH .
\label{a1}
\end{equation}
Keeping J fixed, the differential change of F due to small changes dT and dH is:
\begin{equation}\tag{A2}
dF (H, J, T) = - MdH -SdT .
\label{a2}
\end{equation}
T and H are the macroscopic parameters which can affect F; a variation of the microscopic parameter J shows up in a variation of the ratio (T/J), which is the quantity that appears in F. 

Our analysis of the possible states of the plant relies on the competition between H and K(=T/J): whereas H tends to increase the absolute value of Mo, increasing T tends to increase the entropy S which results in decreasing Mo (the same happens when we decrease the contact J). The actual value of M is the optimal value obtained by those competing effects, i.e., the value which minimizes F.

\end{document}